\documentclass[reprint, superscriptaddress, aps, prb]{revtex4-2}
\usepackage[dvipsnames,svgnames,x11names]{xcolor}
\usepackage{comment}
\usepackage{amsmath,amssymb,amsfonts}
\usepackage{physics}
\usepackage{mathtools}
\usepackage{braket}
\usepackage{graphicx}% Include figure files
\usepackage{dcolumn}% Align table columns on decimal point
\usepackage{bm}% bold math
\usepackage{hyperref}% add hypertext capabilities
\usepackage[utf8]{inputenc}
\usepackage[american,british]{babel}
\usepackage[T1]{fontenc}
\usepackage{bm}
\usepackage{bbold}     
\usepackage{color}
\usepackage{soul} 
\usepackage{orcidlink}

\renewcommand{\rm}[1]{\textrm{#1}}

\newcommand{\mbf}[1]{\mathbf{#1}}
\newcommand{\avg}[1]{\langle #1\rangle}
\renewcommand{\dd}{\mathrm d}
\newcommand{\ii}{\mathrm i}

\newcommand{\tth}{\tau_{\rm{th}}}
\newcommand{\td}{\tau_{\rm{d}}}
\newcommand{\ts}{\tau_{\rm{s}}}
\newcommand{\kk}{\mathbf{k}}
\newcommand{\cF}{\mathcal{F}}

\begin{document}

\title{Beyond the Kubo formula: unification of adiabatic and isothermal response in quantum systems}
\author{Giuliano Chiriacò~\orcidlink{0000-0002-3906-4437}}
\email{giuliano.chiriaco@dfa.unict.it}
\affiliation{Dipartimento di Fisica e Astronomia, Università di Catania, Via Santa Sofia 64, 95123, Catania, Italy }
\author{Vasco Cavina}
%\affiliation{Scuola Normale Superiore and Istituto di Scienze Umane ed Esistenziali}
\affiliation{NEST, Scuola Normale Superiore, Piazza dei Cavalieri 7, I-56126 Pisa, Italy}

\begin{abstract}
The static limit of the Kubo susceptibility describes the adiabatic response of an isolated quantum system, in which energy-level occupations remain unchanged, and generally differs from the isothermal susceptibility, where these occupations readjust to thermal equilibrium.
In this work, we introduce a single dynamical framework in which the adiabatic and isothermal responses arise naturally as two distinct limiting regimes. This can be done by considering explicitly an open system version of the Kubo approach in which the dissipation is governed by a typical thermalization rate of the system. By exchanging the order of the static limit and of the vanishing thermalization rate limit, the protocol yields either the adiabatic or isothermal response. This protocol has a solid physical motivation in terms of the thermalization dynamics and is valid for generic systems. We apply it to the case of the magnetic response of the electron gas and reproduce the expected results.
\end{abstract}

\maketitle

Linear response theory is the main bridge between the microscopic description of a many-body system and the quantities measured in experiments. The Kubo formula expresses the response of an observable to a weak external perturbation in terms of a retarded correlation function of the unperturbed system~\cite{Kubo,kuboStatisticalPhysicsII1991,giulianiQuantumTheoryElectron2005}, and allows the calculation of essentially all response coefficients. The formula is derived for a closed system evolving unitarily under the perturbed Hamiltonian, and static susceptibilities are obtained as the zero-frequency limit of the corresponding dynamical response functions.

It was recognized already by Kubo that this zero-frequency limit, also called adiabatic or isolated susceptibility, does not coincide in general with the isothermal susceptibility, defined as the second derivative of the equilibrium free energy~\cite{Kubo,wilcoxBoundsIsothermalAdiabatic1968,suzukiErgodicityConstantsMotion1971,falkLowerBoundIsothermal1968,mazurNonergodicityPhaseFunctions1969}. A textbook example is the spin susceptibility of a free electron gas: since the total spin $\hat S_z$ commutes with the Hamiltonian, the Kubo response to a uniform static magnetic field vanishes identically, while the static spin susceptibility of a normal metal is given by the (isothermal) Pauli result~\cite{giulianiQuantumTheoryElectron2005}. In practice, the textbook prescription to obtain the static isothermal response is through a specific sequence of limits: the thermodynamic limit is taken before the adiabatic switching of the perturbation is removed, or the static limit $\omega\to0$ is taken before the long-wavelength limit $\mbf q\to0$ \cite{pinesTheoryQuantumLiquids1966,zubarevNonequilibriumStatisticalThermodynamics1974,giulianiQuantumTheoryElectron2005,merminLindhardDielectricFunction1970,ilievskiThermodyamicBoundsDrude2013,chibaAnomalousBehaviorMagnetic2020,floerchingerResponseTheoryQuantum2026}.

The problem has acquired new relevance in recent years. These order-of-limits prescriptions rely on the thermodynamic limit and translational invariance, so that they do not apply to finite or inhomogeneous systems, such as quantum dots, qubits, or trapped atomic gases, and they do not identify which physical timescale controls the crossover between the two responses. Moreover, the Kubo formula contains a single timescale, set by the frequency of the perturbation, and has no knowledge of the thermalization processes that determine whether the response of a system is adiabatic or isothermal.

Indeed, there has recently been great interest regarding the thermalization and relaxation of many-body quantum systems. Several works have studied whether and how isolated systems thermalize \cite{Deutsch91,Srednicki94,rigolThermalizationItsMechanism2008,Alessio16,rigolRelaxationCompletelyIntegrable2007,polkovnikovColloquiumNonequilibriumDynamics2011,gogolinEquilibrationThermalisationEmergence2016,moriThermalizationPrethermalizationIsolated2018}, how relaxation affects their properties and whether new physical phases may emerge as a result \cite{fazioManyBodyOpenQuantum2024,LiFisher2018,Noel21,GullansHuse2020a,sierantDissipativeFloquetDynamics2022a,chiriacoDiagrammaticMethodManybody2023a,Turkeshi2021:Ising,muzziEntanglementEnhancementInduced2025,xing2025entanglementtransitionunitarysystembath,basovElectrodynamicsCorrelatedElectron2011,Aver:opt-rev,mitranoPossibleLightinducedSuperconductivity2016,mciverLightinducedAnomalousHall2020,chiriacoNegativeAbsoluteConductivity2020b,kogar_light-induced_2020,zong_evidence_2019,Mengkun:CRO,Maeno:mainp,chiriacoVoltageinducedMetalinsulatorTransition2018b,jarc_cavity-mediated_2023,chiriacoThermalPurcellEffect2024}, and how thermalization occurs for driven systems coupled to external environments\cite{quanQuantumThermodynamicCycles2007,cavinaSlowDynamicsThermodynamics2017,scandiThermodynamicLengthOpen2019,yamaguchiMarkovianQuantumMaster2017,dechiaraReconciliationQuantumLocal2018,e.chiattoLocalVsGlobal2026}. More recent works have studied linear response to a perturbation of the dissipative processes within a Lindblad setting \cite{camposvenutiAdiabaticityOpenQuantum2016,camposvenutiDynamicalResponseTheory2016,panNonHermitianLinearResponse2020,blairNonequilibriumQuantumProbing2024,konopikQuantumResponseTheory2019,mehboudiLinearResponseTheory2019,levyResponseTheoryNonequilibrium2021,talkingtonLinearNonlinearResponse2024,albertGeometryResponseLindbladians2016,banLinearResponseTheory2017,albashQuantumAdiabaticMarkovian2012,wangNonadiabaticEvolutionThermodynamics2021,albashAdiabaticQuantumComputation2018}.

Experimental advances make it possible to keep ultracold atomic gases isolated from the environment over long times~\cite{blochManybodyPhysicsUltracold2008}, and to observe directly the unitary dynamics of many-body systems and their relaxation~\cite{kinoshitaQuantumNewtonCradle2006,gringRelaxationPrethermalizationIsolated2012,trotzkyProbingRelaxationEquilibrium2012}. On the other hand, in solid-state quantum devices the frequency of the probe can be comparable to the relaxation rates of the system. An example is the response of quantum dots and circuits, which displays an adiabatic contribution and a thermalization contribution ~\cite{mizutaQuantumTunnelingCapacitance2017,periBeyondadiabaticQuantumAdmittance2024,periUnifiedLinearResponse2024,kitsenkoReflectionsQuantumReflectometry2026,kuhnLindbladTheoryLinear2026}. 

\begin{figure}
    \centering
\includegraphics[width=\linewidth]{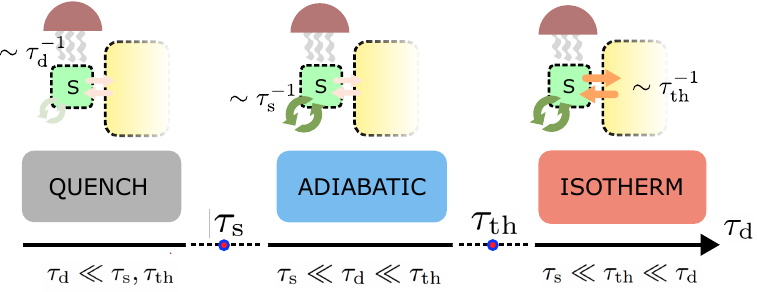}
    \caption{Pictorial summary of the three regimes considered in the paper. In the quench regime, the driving speed (proportional to $\tau_d^{-1}$) is much faster than the rates associated to thermalization $\tth^{-1}$ and internal dynamics $\tau_s^{-1}$ of the system. In the adiabatic regime, the internal dynamics becomes the faster one, while the standard prescription of the adiabatic theorem can be applied, neglecting all effects due to thermalization. Finally, in the isotherm, thermalization becomes faster than the driving and emerges as the relevant time scale.   
    }
    \label{fig:Figure1}
\end{figure}

Thus, it is of great interest to both theory and experiments to study the response of a system in the presence of thermalization processes, and develop a general framework that makes the role of thermalization explicit and that yields both the adiabatic and the isothermal response within the same calculation.

In this work we address this problem. We identify the dynamical regimes of a quantum system subject to a time-dependent perturbation by comparing the perturbation timescale  with the intrinsic timescale of the system and with its thermalization time, and we derive the corresponding static susceptibilities. We show that the zero-frequency limit of the Kubo response yields the adiabatic susceptibility, in which the populations of the instantaneous eigenstates are frozen, and which coincides with the isolated susceptibility of Refs.~\cite{Kubo,wilcoxBoundsIsothermalAdiabatic1968}, and we show that it generally differs from the isothermal response. For the self-response of an operator, we also find a hierarchy between susceptibilities. 

We then generalize the Kubo formula by explicitly including the thermalization rate $\kappa$ in the response function $\chi(\omega,\kappa)$. Within this unified framework, the adiabatic and isothermal susceptibilities are naturally obtained by taking the limits $\omega\to0$ and $\kappa\to0$ in the appropriate order. Our framework is valid for generic systems, and also interpolates between adiabatic and isothermal responses, allowing to study intermediate regimes. We apply this protocol to the paramagnetic response of a free electron gas and show that it correctly yields a vanishing adiabatic response and the Pauli isothermal susceptibility.

\emph{The system}. -- We consider a many-body quantum system described by the Hamiltonian $\hat H_0$, with eigenstates $\ket{n}$ and eigenvalues $E_n$, namely $\hat{H}_0\ket{n}=E_n\ket{n}$. The system is subject to a time dependent perturbation $\lambda(t)$ that couples to an operator $\hat B$
\begin{equation}
\hat{H}_\lambda=\hat{H}_0+\lambda(t)\hat{B}; \qquad \hat{H}_\lambda\ket{n(\lambda)}=E_n(\lambda)\ket{n(\lambda)},
\label{Eq:Hamiltonian}
\end{equation}
where $\ket{n(\lambda)}$ and $E_{n}(\lambda)$ are the instantaneous eigenstates and the respective eigenvalues of $\hat H_\lambda$.
We are interested in investigating the response of the system  to the external perturbation within the linear regime, by studying the expectation value of an operator $\hat A$. This expectation value is given, under the assumption of weak perturbations, by the Kubo formula \cite{giulianiQuantumTheoryElectron2005}
\begin{equation}\label{Eq:Avalue}
A(t)=\operatorname{Tr}[\rho(t)\hat A]=A_0+\int_{-\infty}^{\infty}\dd t'\,\chi_{AB}(t-t')\lambda(t').
\end{equation}
where $A_0=\text{Tr}(\rho_0\hat A)$ is the expectation value in the initial state and $\chi_{AB}(t)\equiv-i\theta(t)\avg{[\hat A(t),\hat B]}$ is the retarded component of the susceptibility. 

We assume that the system is initially in a thermal state $\rho_0=\frac1Z\sum_ne^{-\beta E_n}\ketbra{n}{n}$ with $\beta$ the inverse temperature and $Z=\sum_ne^{-\beta E_n}=\text{Tr}(e^{-\beta\hat H_0})$ the partition function. The perturbation is ramped up from $0$ to $\lambda$ over a typical time $\td$: depending on how fast the perturbation changes, the system evolves according to different dynamical regimes, see Fig. \ref{fig:Figure1}, which are associated to different values of the final density matrix $\rho'$.

\textit{(i) Quench regime. -- } If $\lambda$ varies very fast ($\td\rightarrow0$), the system does not have the time to react to the perturbation and the density matrix does not change
$\rho'=%\rho_Q
\rho_0$.

\textit{(ii) Quantum Adiabatic regime. -- } If $\lambda$ varies slowly enough, compared to the typical timescale $\ts$ of the system, then the dynamics satisfies the adiabatic theorem. The timescale $\ts$ is typically determined by the energy gaps of the unperturbed Hamiltonian. In this regime, a system prepared in $\ket{n}$ remains in the instantaneous eigenstate $\ket{n(\lambda)}$ at each step of the transformation, while the populations are left unchanged, so that the final density matrix is
\begin{equation}\label{Eq:rhoS}
\rho'=\rho_S \equiv \frac1Z\sum_ne^{-\beta E_n}\ketbra{n(\lambda)}{n(\lambda)}.
\end{equation}

Note that the final state is not thermal; in fact, although the density matrix is diagonal, the populations $p_n=\rho_{nn}$ do not satisfy a detailed-balance relation with respect to the new energies $p_{n}/p_{n'}=e^{-\beta E_n}/e^{-\beta E_{n'}}\neq e^{-\beta E_n(\lambda)}/e^{-\beta E_{n'}(\lambda)}$.

\textit{(iii) Isothermal regime. --} The quench and adiabatic regimes emerge in closed systems, where no thermalization process can change the populations of the energy levels. However, a generic many-body system interacts with an external environment that acts as a thermal bath \cite{breuerTheoryOpenQuantum2007}. In our case, we assume that the effect of the environment is to make the system relax to a thermal state with inverse temperature $\beta^{-1}$, on a typical timescale $\tth$. When the perturbation occurs on timescales longer than the relaxation time (i.e. $\tau_\rm d\gg\tau_{\rm{th}}$), the system is in the instantaneous thermal state
\begin{equation}\label{Eq:rhoT}
\rho'=\rho_T \equiv \frac1{Z(\lambda)}\sum_ne^{-\beta E_n(\lambda)}\ketbra{n(\lambda)}{n(\lambda)},
\end{equation}
where $Z(\lambda)=\sum_ne^{-\beta E_n(\lambda)}$. We note that $\rho_S$ and $\rho_T$ differ for the value of the populations, and that both the quantum adiabatic and isothermal regimes correspond to quasistatic transformations, which occur on a timescale much longer than $\ts$.
Also note that the quantum-adiabatic regime considered here differs from classical thermodynamic adiabaticity, which assumes internal thermalization during a quasistatic evolution. In isolated many-body systems, such thermalization can occur in systems satisfying the eigenstate thermalization hypothesis \cite{Deutsch91,Srednicki94,rigolThermalizationItsMechanism2008,Alessio16}. However, the effective temperature is determined by the conservation of total entropy and may vary with external parameters. The resulting response therefore generally differs from the isothermal one~\cite{suzukiErgodicityConstantsMotion1971,chibaAnomalousBehaviorMagnetic2020,floerchingerResponseTheoryQuantum2026}. A detailed investigation of this regime is left for future work.

\emph{Susceptibilities}. -- We now want to explore the susceptibility in the various regimes sketched in Fig. \ref{fig:Figure1}. We use the superscripts $Q,S,T$ to denote quench, adiabatic and isothermal quantities, respectively, while we use ${\rm qs}$ for denoting a generic quasistatic process.

In the quench regime the susceptibility vanishes because the state does not change, $\chi^Q=0$. This is compatible with the $\omega\rightarrow\infty$ limit of the Kubo susceptibility.%, which vanishes.  
To analyze the adiabatic and isothermal regimes, we write Eq. \eqref{Eq:Avalue} as
\begin{equation} \label{eq:neweq}
A(t)-A_0=\int\frac{\dd\omega}{2\pi}e^{-\ii\omega t}A(\omega)=\int\frac{\dd\omega}{2\pi}e^{-\ii\omega t}\chi_{AB}(\omega)\lambda(\omega)
\end{equation}
using the Fourier transforms of $A$, $\chi_{AB}$ and $\lambda$.

Within these regimes, the perturbation is quasistatic, i.e. $\lambda$ changes very slowly from zero to its final value, and $\lambda(\omega)$ is peaked near $\omega=0$, with width $\omega\lesssim1/\td$, so that $\chi_{AB}(\omega)\simeq\chi_{AB}(\omega\to0)$.
We can take this limit as a definition of the quasistatic susceptibility, so that, after replacing in Eq.\eqref{eq:neweq}, we obtain
\begin{gather}\label{Eq:Aqs}
A^{\rm qs}(t)-A_0\simeq\chi_{AB}(\omega\to0)\lambda(t)=\chi_{AB}^{\rm qs}\lambda(t).
\end{gather}

The limitation of the Kubo formalism is that the above description should be universally valid for quasistatic processes and thus encompass both the adiabatic and isothermal regimes.
In contrast, adiabatic and isothermal susceptibilities can in general differ. 
In fact, from Eqs. \eqref{Eq:Avalue}-\eqref{Eq:rhoT} we write
\begin{gather}\label{Eq:ASAT}
    A^S-A_0=\frac1Z\sum_ne^{-\beta E_n}A_{nn}(\lambda);\\
    A^T-A_0=\frac1{Z(\lambda)}\sum_ne^{-\beta E_n(\lambda)}A_{nn}(\lambda),
\end{gather}
and derive the respective susceptibilities \cite{giulianiQuantumTheoryElectron2005,wilcoxBoundsIsothermalAdiabatic1968,suzukiErgodicityConstantsMotion1971}
\begin{gather}\label{Eq:LehmannS}
\chi_{AB}^S=\frac1{Z}\sum_{m\neq n}\frac{e^{-\beta E_m}-e^{-\beta E_n}}{E_m-E_n}A_{mn}B_{nm};\\
\label{Eq:LehmannT}\chi_{AB}^T=\chi_{AB}^S-\beta(\avg{\hat A_0\hat B_0}_0-\avg{\hat A_0}_0\avg{\hat B_0}_0),
\end{gather}
where $A_{mn}(\lambda)\equiv \braket{m(\lambda)|\hat A|n(\lambda)}$ and $\hat A_0\equiv \sum_nA_{nn}\ketbra{n}{n}$ is the diagonal part of $\hat A$ in the eigenbasis of $\hat H_0$ and $\avg{\cdot}_0=\text{Tr}(\rho_0\,\cdot)$ is the expectation value on the unperturbed state.

Comparing with the static limit of the Kubo formula, we see that $\chi_{AB}^S=\lim_{\omega\rightarrow0}\chi_{AB}(\omega)$. However, in general $\chi^S\neq\chi^T$, and even though it is a quasistatic susceptibility, $\chi_{AB}^T$ \emph{cannot} be obtained as $\chi_{AB}(\omega\to0)$, as it was already pointed out in Refs.~\cite{Kubo,falkLowerBoundIsothermal1968,wilcoxBoundsIsothermalAdiabatic1968,mazurNonergodicityPhaseFunctions1969,suzukiErgodicityConstantsMotion1971}.

This apparent paradox is a limitation of the Kubo formulation, which is derived for a closed system and does not include any thermalization process, so that the ${\omega\to0}$ limit gives a quasistatic response in a context in which the evolution is unitary, i.e. the adiabatic response.
 
To treat the isothermal response in the standard Kubo approach, we have to introduce a suitable model of thermalizaton, associated to a typical time $\tth$, or, equivalently, a typical thermalization rate $\kappa = \tth^{-1}$  (see Fig. \ref{fig:Figure1}). We thus generalize the Kubo response by taking into account the full open evolution, containing the unitary part (controlled by $\omega$) and the dissipative part (controlled by $\kappa$), leading to a susceptibility of the form $\chi_{AB}(\omega,\kappa)$. The quasistatic regime is then obtained by sending both $\kappa$ and $\omega$ to zero. However, the two limits do not commute and correspond to the different physical scenarios we are interested in:
\begin{equation}\label{Eq:Limits}
\boxed{
\begin{gathered}
\quad\chi_{AB}^S=\lim_{\omega\to0}\lim_{\kappa\to0}\chi_{AB}(\omega,\kappa),\quad\\
\quad\chi_{AB}^T=\lim_{\kappa\to0}\lim_{\omega\to0}\chi_{AB}(\omega,\kappa).\quad
\end{gathered}}
\end{equation}

We stress that, not only the adiabatic and isothermal susceptibilities can be naturally obtained via the limits in Eq. \eqref{Eq:Limits}, but that the general formula for $\chi_{AB}(\omega,\kappa)$ is also valid in intermediate regimes interpolating between adiabatic and isothermal response. Our result is thus relevant to many experimental platforms, whose response ranges from isothermal (most solid-state systems), to adiabatic (atomic and optical systems \cite{blochManybodyPhysicsUltracold2008}), to regimes where $\omega$ is comparable to $\kappa$ (e.g. quantum devices probed at microwave frequencies \cite{mizutaQuantumTunnelingCapacitance2017}).

A physically relevant case, that applies for instance to the density-density response, the electrical conductivity, the magnetic susceptibility, is the one in which $\hat B\propto\hat A$, so that the response is proportional to $\chi_{AA}$. From Eq. \eqref{Eq:LehmannS} we show that the adiabatic susceptibility is always negative $\chi_{AA}^S=\frac1Z\sum_{m\neq n}\frac{e^{-\beta E_m}-e^{-\beta E_n}}{E_m-E_n}|A_{mn}|^2\leq0$, because $\frac{e^{-\beta E_m}-e^{-\beta E_n}}{E_m-E_n}\leq0$.

Moreover, $\chi_{AA}^T-\chi_{AA}^S=-\beta(\avg{\hat A_0^2}_0-\avg{\hat A_0}_0^2)\leq0$, since the variance of $\hat A_0$ is always positive, so that $\chi_{AA}^T\leq\chi_{AA}^S$. The susceptibilities then satisfy a hierarchy:
\begin{equation}\label{Eq:hierarchy}
\chi_{AA}^T\leq\chi_{AA}^S\leq\chi_{AA}^Q=0
\end{equation}
which is consistent with the bounds between adiabatic and isothermal susceptibilities derived in Refs.~\cite{falkLowerBoundIsothermal1968,wilcoxBoundsIsothermalAdiabatic1968,suzukiErgodicityConstantsMotion1971} and with considerations on energy balance and on the work done on the system by the perturbation \cite{giulianiQuantumTheoryElectron2005}.

\emph{Application to magnetic response}. -- We now consider a free electron gas subject to a static magnetic field $B$, a textbook case in which the adiabatic and isothermal magnetic susceptibilities differ, and use it to demonstrate the protocol of Eq. \eqref{Eq:Limits}. We start from the Hamiltonian:
\begin{equation}\label{Eq:ElectronGas}
\begin{gathered}
\hat H=\int\dd\mbf r\sum_\sigma\hat\psi_\sigma^\dagger(\mbf r)\left(-\frac{\nabla^2}{2m}-\mu\right)\hat\psi_\sigma(\mbf r)-\mu_BB\hat S_z,
\end{gathered}
\end{equation}
where $\mu_B$ is the Bohr magneton, $\hat S_z=\int\dd\mbf  r[\hat n_\uparrow(\mbf r)-\hat n_\downarrow(\mbf r)]$ is the total spin operator, $\hat\psi_\sigma(\mbf r)$ is the electron field operator at position $\mbf r$ and spin $\sigma$, and $\hat n_\sigma(\mbf r)=\psi^\dagger_\sigma(\mbf r)\hat\psi_\sigma(\mbf r)$. We assume the temperature to be low, so that the chemical potential $\mu$ coincides with the Fermi energy $\varepsilon_F$.

\begin{figure}
\includegraphics[width=0.95\linewidth]{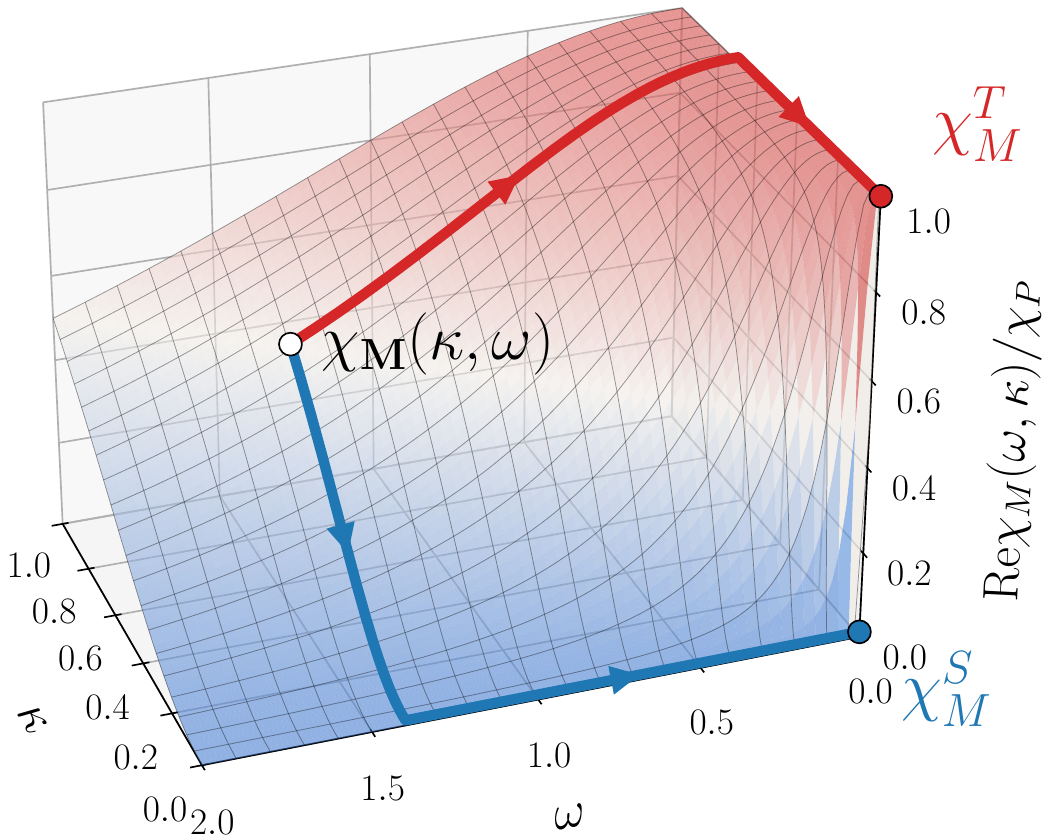}
    \caption{Surface plot of the real part of the generalized susceptibility $\chi_M(\omega,\kappa)$ normalized to $\chi_P$ in the zero temperature limit, as function of $\omega$ and $\kappa$. The two arrows show the different limits ordering of Eq. \eqref{Eq:Limits}, which lead to thw adiabatic (blue) and isothermal (red) response.}
    \label{fig:Figure2}
\end{figure}

The paramagnetic susceptibility is given by
\begin{equation}
\begin{gathered}
\chi_M(\omega) \equiv \frac1{\mathcal{V}}\frac{\partial\avg{\mu_B\hat S_z}}{\partial B}=-\frac{\mu_B^2}{\mathcal V}\chi_{S_zS_z}(\omega),\\
\chi_{S_zS_z}(t,t')=-\ii\theta(t-t')\avg{[\hat S_z(t),\hat S_z(t')]},
\end{gathered}
\end{equation}
where $\mathcal{V}$ is the volume of the system \footnote{Note that the static spin-spin susceptivility is always negative in accordance with Eq. \eqref{Eq:hierarchy}, while the paramagnetic susceptibility is positive.}. The expression for $\chi_M$ is proved in Appendix \ref{App:Magnetic}.
In the absence of any thermalizing process and for a static magnetic field, $\hat S_z$ commutes with $\hat H$ and is a constant of motion, so that $[\hat S_z(t),\hat S_z(t')]=0$, and we obtain
\begin{equation}
\chi_{S_zS_z}=0\,\,\,\Rightarrow\,\,\,\chi_M^S=0.
\end{equation}

In the isothermal regime we expect to retrieve the standard Pauli susceptibility  \cite{giulianiQuantumTheoryElectron2005} \begin{equation}\chi_M^T=\chi_P=\mu_B^2\frac1{\mathcal{V}}\sum_{\kk\sigma}\left(-\frac{\partial f}{\partial\varepsilon_\kk}\right),   
\end{equation}

with $\varepsilon_{\mbf k}=\frac{\mbf k^2}{2m}-\mu$ and $f(\varepsilon)=(e^{\beta\varepsilon}+1)^{-1}$ the Fermi-Dirac distribution function. At zero temperature $\chi_P=\mu_B^2\mathcal{N}(\varepsilon_F)$, with $\mathcal{N}(\varepsilon)\equiv\frac1{\mathcal{V}}\sum_{\mbf k\sigma}\delta(\varepsilon-\mbf k^2/2m)$ the density of states. 
We now want to show that the values of $\chi_M^S$ and $\chi_M^T$ can be obtained within the same framework, by including thermalization (with rate $\kappa$), calculating $\chi_M(\omega,\kappa)$ and taking the limits $\omega,\kappa\rightarrow0$ in the appropriate order.

Here we report the main results, and carry out the detailed calculations in Appendix \ref{App:Magnetic}. We start from the expression for the magnetic susceptibility
\begin{equation}\label{Eq:chiM}
\begin{gathered}
\chi_M(\omega)=-\mu_B^2[\chi_{\uparrow\uparrow}+\chi_{\downarrow\downarrow}-\chi_{\uparrow\downarrow}-\chi_{\downarrow\uparrow}](\mbf{q}=0,\omega)\\
\chi_{\sigma\sigma'}(\mbf r,t;\mbf r',t')=-\ii\theta(t-t')\avg{[\hat n_\sigma(\mbf r,t),\hat n_{\sigma'}(\mbf r',t')]}.
\end{gathered}
\end{equation}

The spin resolved density-density susceptibility is calculated using standard Green's functions techniques \cite{Kamenev01052009}
\begin{align}
 \notag   \chi_{\sigma\sigma'}(0,\omega)=&\frac{-\ii}{2\mathcal V}\sum_{\mbf k}\int\frac{\dd\varepsilon}{2\pi}\big[G_{\sigma\sigma'}^R(\mbf{k},\varepsilon)G_{\sigma'\sigma}^K(\kk,\varepsilon-\omega)\\
\label{Eq:chiSS}    +&G_{\sigma\sigma'}^K(\mbf{k},\varepsilon)G_{\sigma'\sigma}^A(\mbf{\mbf{k}},\varepsilon-\omega)\big],
\end{align}
where $G^{R/A}_{\sigma\sigma'}\equiv\mp\ii\theta(\pm(t-t'))\avg{[\hat\psi_\sigma(\mbf r,t),\hat\psi^\dagger_{\sigma'}(\mbf r',t')]}$ is the retarded/advanced component of the Green's function, and the Keldysh component $G^K$ is given by $G_{\sigma\sigma'}^K(\mbf{k},\varepsilon)=[1-2f(\varepsilon)][G_{\sigma\sigma'}^R(\mbf{k},\varepsilon)-G_{\sigma\sigma'}^A(\mbf{k},\varepsilon)]$. In order to calculate $\chi_M(\omega,\kappa)$ we just need to calculate $G^{R/A/K}$ in the presence of a suitable thermalization process.

\textit{External thermalization}. -- We now assume that the system interacts with a fermionic bath at temperature $\beta^{-1}$ through a hopping interaction, which allows the exchange of electrons with the environment at rate $\kappa$. We trace out the bath and find that the Green's functions are $G_{\sigma\sigma'}^{R/A/K}=\delta_{\sigma\sigma'}G^{R/A/K}$ \cite{Kamenev01052009,siebererKeldyshFieldTheory2016a,thompsonFieldTheoryManybody2023}, with:
\begin{equation}\label{Eq:GreenKeldyshLindblad}
G^{R/A}=\frac{1}{\varepsilon-\varepsilon_{\mbf k}\pm\ii\kappa};\qquad
G^K=\frac{-2\ii\kappa[1-2f(\varepsilon)]}{(\varepsilon-\varepsilon_{\mbf k})^2+\kappa^2}.
\end{equation}

We substitute into Eq. \eqref{Eq:chiSS} and carry out the calculations, see Appendix \ref{App:External bath}. At zero temperature we obtain an explicit expression for $\chi_M(\omega,\kappa)$:
\begin{equation}\label{Eq:chiMT0}
\chi_M(\omega,\kappa)=\chi_P\frac{2\kappa}{2\kappa-i\omega}.
\end{equation}

From Eq. \eqref{Eq:chiMT0} it is clear that the two limits do not commute and that they yield the expected results: $\chi_M^S=\lim_{\omega\rightarrow0}\lim_{\kappa\rightarrow0}\chi_M=0$ and $\chi_M^T=\lim_{\kappa\rightarrow0}\lim_{\omega\rightarrow0}\chi_M=\chi_P$. The real part of $\chi_M(\omega,\kappa)$ is plotted in Fig. \ref{fig:Figure2}.

The same result is obtained at generic temperatures, for which we find $\chi_M(\omega,\kappa=0)=0$, so that $\chi_M^S=0$, and $\chi_M^T=\mu_B^2\frac1{\mathcal{V}}\sum_{\mbf k,\sigma}\left(-\frac{\partial f}{\partial\varepsilon_{\mbf k}}\right)$, which coincides with the Pauli susceptibility at finite temperatures. 

Incidentally, we also obtained a generalized susceptibility expression [Eq. \eqref{Eq:chiMT0}] that interpolates between the adiabatic and isothermal regimes, making it applicable to scenarios that exhibit finite values of $\omega$ and $\kappa$ \cite{mizutaQuantumTunnelingCapacitance2017}.

Note that the origin of thermalization mechanism is not important, provided that the interaction with the bath generates a self-energy with a non-zero imaginary part, which in our model we assume to be constant, i.e. $\kappa=-\text{Im}\Sigma$. Other types of interaction can be used to calculate its corresponding self-energy and thus $\chi(\omega,\kappa)$. For example, we may imagine thermalization arising from electron-electron, electron-phonon or electron-impurity interactions. The only requirement is that $\kappa$ respects the conservation laws of the system
\cite{merminLindhardDielectricFunction1970}.

A subtlety arises at finite momentum~\cite{giulianiQuantumTheoryElectron2005,chibaAnomalousBehaviorMagnetic2020}: calculating $\chi_M(\mbf q,\omega,\kappa)$ from $\chi_{\sigma\sigma'}(\mbf q,\omega)$, the limits $\omega\rightarrow0$ and $\kappa\rightarrow0$ commute, yielding
\begin{equation}
\chi_M^S(\mbf q)=\chi_M^T(\mbf q)=\frac{\mu_B^2}{\mathcal V}\sum_{\mbf k\sigma}\frac{f(\varepsilon_{\mbf k})-f(\varepsilon_{\mbf k-\mbf q})}{\varepsilon_{\mbf k-\mbf q}-\varepsilon_{\mbf k}+\ii\eta}.
\end{equation}

The subsequent limit $\mbf q\rightarrow0$ gives the static isothermal response $\chi_P=\frac{\mu_B^2}{\mathcal V}\sum_{\mbf k\sigma}(-\partial_{\varepsilon_{\mbf k}}f)$. Thus, $\chi_M(\mbf q)$ is discontinuous at $\mbf q=0$:
$\lim_{\mbf q\rightarrow0}\chi_M(\mbf q)=\chi_M^T\neq\chi_M^S=\chi_M(\mbf q=0)$.

The equality $\chi_M^T(\mbf q)=\chi_M^S(\mbf q)$ at $\mbf q\neq0$ follows from $\chi_M(\mbf q)\sim\avg{\hat n_\mbf q\hat n_{-\mbf q}}$, where $\hat n_\mbf q\sim\sum_{\mbf k}\hat c^\dagger_\mbf k\hat c_{\mbf k-\mbf q}$. By Eq.~\eqref{Eq:ASAT}, their difference depends on expectation values of $\hat n_\mbf q$ onto eigenstates of $\hat H_0$. Translational invariance allows these states to have definite momentum, so these expectation values vanish for $\mbf q\neq0$, since $\hat n_\mbf q$ changes momentum. This explains the textbook prescription of taking $\omega\rightarrow0$ before $\mbf q\rightarrow0$ to obtain the static isothermal response~\cite{pinesTheoryQuantumLiquids1966,zubarevNonequilibriumStatisticalThermodynamics1974,giulianiQuantumTheoryElectron2005}. Its validity is therefore a lucky accident, contingent on translational symmetry and the symmetry-breaking response operator, rather than a fundamental principle. In contrast, our physically motivated protocol applies to any system.

In appendix \ref{App:spin flip}, we study a similar setup where no bath is present but the Hamiltonian contains a spin-flip term. Even if this term does not induce a thermalization process, it breaks the spin-symmetry that makes $\hat S_z$ a constant of motion, and its introduction allows to correctly calculate $\chi_M^S$ and $\chi_M^T$. We study it because it displays a similar phenomenology as adiabatic quantum annealing in the transverse-field Ising model ~\cite{kadowakiQuantumAnnealingTransverse1998,dasColloquiumQuantumAnnealing2008,albashAdiabaticQuantumComputation2018,shevchenkoLandauZenerStuckelbergInterferometry2010,dicksonThermallyAssistedQuantum2013,aminSearchingQuantumSpeedup2015,marshallPowerPausingAdvancing2019}. In Appendix \ref{App:class}, we also work out a simple classical problem which displays different adiabatic and isothermal responses.

\emph{Conclusions and outlook}. -- We have studied the response of a quantum system subject to an external perturbation. We have derived a generalized susceptibility that incorporates both driving ($\td$) and thermalization ($\tth$) time scales, extending the Kubo approach to recover both isothermal and adiabatic responses within a unified framework. Without explicit thermalization, the Kubo formula yields only the adiabatic response. Our approach recovers both responses in the quasistatic limit: taking $\td\rightarrow \infty$ before $\tau_{\mathrm{th}}\rightarrow \infty$ yields the isothermal response, while reversing the order yields the adiabatic response.

Our approach is based on clean physical motivations and corrects standard textbook prescriptions that fail for finite systems or in the absence of translational invariance. This is particularly relevant when thermalization cannot be suppressed, as in most solid-state systems, and for finite-size systems, such as trapped atoms and qubits. Applied to the magnetic susceptibility of a free electron gas, our protocol reproduces the expected results within a framework applicable to any system.

Several questions warrant further investigation. The finite-momentum magnetic response is discontinuous at $\mbf q=0$: both adiabatic and isothermal susceptibilities at $\mbf q\neq0$ approach the uniform isothermal response, rather than the uniform adiabatic response. The origin of this connection deserves further study. Their distinct responses could also be exploited in quantum devices~\cite{mizutaQuantumTunnelingCapacitance2017,periBeyondadiabaticQuantumAdmittance2024,periUnifiedLinearResponse2024,kitsenkoReflectionsQuantumReflectometry2026,kuhnLindbladTheoryLinear2026}, including sensors of perturbation frequency or strength. Another direction is to calculate susceptibility corrections in intermediate regimes using adiabatic and quasistatic expansions~\cite{cavinaSlowDynamicsThermodynamics2017,scandiThermodynamicLengthOpen2019, cavina2021maximum}.

\textit{Acknowledgements --} We thank Gian Marcello Andolina for insightful discussions. We thank R. Fazio and L. Mazza, who years ago assigned us a statistical mechanics exam that inspired this work. G.C. is supported by MUR under project PN-RIC 21-27, action 1.1.2 Q-Sud. G.C. acknowledges the CINECA award under the ISCRA initiative, for the availability of high performance computing resources and support. AI tools (ChatGPT and Claude) were used to proof-read the manuscript.

\bibliography{biblio}
\bibliographystyle{apsrev}

\newpage

\appendix
\onecolumngrid

\section{Susceptibility via the free energy}\label{App:Susceptibility} 

In this section we express the adiabatic and thermal susceptibility in terms of the free energy of the system, $F\equiv - \frac1\beta\ln(\sum_ne^{-\beta E_n})$. The expectation values of Eq. \eqref{Eq:ASAT} can be obtained introducing an additional source term in the Hamiltonian $\hat H(\lambda_A,\lambda)\equiv\hat H_0+\lambda_A\hat A+\lambda\hat B$, and writing them as derivatives of the free energy $F(\lambda_A,\lambda)\equiv-\frac1\beta \ln\text{Tr}(e^{-\beta\hat H(\lambda_A,\lambda)})$. We use that at first order perturbation theory in $\lambda_A$ it holds $E_n(\lambda_A,\lambda)=E_n(\lambda)+\lambda_AA_{nn}(\lambda)+\mathcal{O}(\lambda_A^2)$ and write
\begin{gather}
A^T=\frac1{Z(\lambda)}\sum_ne^{-\beta E_n(\lambda)}A_{nn}(\lambda)=-\frac1\beta\frac{\partial}{\partial\lambda_A}\ln(\sum_ne^{-\beta E_n(\lambda_A,\lambda)})\Big|_{\lambda_A=0}=\frac{\partial}{\partial\lambda_A}F(\lambda_A,\lambda)\Big|_{\lambda_A=0}.
\end{gather}

For the adiabatic response we find
\begin{gather}
A^S=\frac{1}{Z_0}\sum_ne^{-\beta E_n}A_{nn}(\lambda)=\frac{1}{Z_0}\sum_ne^{-\beta E_n}\frac{\partial E_n(\lambda_A,\lambda)}{\partial\lambda_A}\Big|_{\lambda_A=0}=\sum_n\frac{\partial}{\partial E_n}F(\lambda_A,\lambda)\Big|_{\lambda_A=0,\lambda=0}\frac{\partial E_n(\lambda_A,\lambda)}{\partial\lambda_A}\Big|_{\lambda_A=0}.
\end{gather}

The susceptibility can then be written as $\chi_{AB}=\frac{\partial A}{\partial\lambda}\Big|_{\lambda=0}$, finding

\begin{gather}
\chi_{AB}^S=\sum_n\frac{\partial F(0,0)}{\partial E_n}\frac{\partial^2E_n}{\partial\lambda_A\partial\lambda};\\
\chi_{AB}^T=\frac{\partial^2F(\lambda_A,\lambda)}{\partial\lambda_A\partial\lambda}\Big|_{\substack{\lambda_A=0\\\lambda=0}}=\sum_n\frac{\partial F(0,0)}{\partial E_n}\frac{\partial^2E_n}{\partial\lambda_A\partial\lambda}+\sum_{nm}\frac{\partial E_m}{\partial\lambda}\frac{\partial^2 F(0,0)}{\partial E_n\partial E_m}\frac{\partial E_n}{\partial\lambda_A}=\chi_{AB}^S+\sum_{nm}\frac{\partial E_m}{\partial\lambda}\frac{\partial^2 F(0,0)}{\partial E_n\partial E_m}\frac{\partial E_n}{\partial\lambda_A}.
\end{gather}

Geometrically, the adiabatic susceptibility is obtained by evaluating the mixed derivative of the tangent-plane approximation to the free energy in the space of energy levels, corresponding to fixed occupation probabilities. The isothermal susceptibility additionally accounts for the curvature of the free energy in this space, which describes the redistribution of thermal populations.

\section{Magnetic susceptibility}\label{App:Magnetic}

In this section we show the detailed steps of the calculations related to the magnetic susceptibility.

We want to calculate the paramagnetic susceptibility of a free electron gas subject to a magnetic field $\bm B(t)=B(t)\hat z$:
\begin{equation}H=\int\dd\mbf r\sum_\sigma\hat\psi_\sigma^\dagger(\mbf r)\left(-\frac{\nabla^2}{2m}-\mu\right)\hat\psi_\sigma(\mbf r)-\mu_BB(t)\hat S_z,\end{equation}
where $\mu_B$ is the Bohr magneton. We have assumed $\hbar=1$ and multiplied the factor $1/2$ coming from the spin definition with the electron giromagnetic factor $g_s=2$, so that the total electronic spin is 
\begin{equation}\hat S_z=\int\dd\mbf r[\hat n_\uparrow(\mbf r)-\hat n_\downarrow(\mbf r)]=\int\dd\mbf r[\hat\psi_\uparrow^\dagger(\mbf r)\hat\psi_\uparrow(\mbf r)-\hat\psi_\downarrow^\dagger(\mbf r)\hat\psi_\downarrow(\mbf r)].\end{equation}

The magnetization per unit volume is
\begin{gather}
M(t)=\frac1{\mathcal{V}}\mu_B\avg{\hat S_z(t)}=\int\dd t'\,\chi_M(t,t')B(t');\qquad\chi_M(t,t')=-\frac{\mu_B^2}{\mathcal V}\chi_{S_zS_z}(t,t');\\
\chi_{S_zS_z}(t,t')=-\ii\theta(t-t')\avg{[\hat S_z(t),\hat S_z(t')]}=-\ii\theta(t-t')\int\dd\mbf r\dd\mbf r'\avg{[\hat n_\uparrow(\mbf r,t)-\hat n_\downarrow(\mbf r,t),\hat n_\uparrow(\mbf r',t')-\hat n_\downarrow(\mbf r',t')]};\\
\chi_{S_zS_z}=\int\dd\mbf r\dd\mbf r'(\chi_{\uparrow\uparrow}-\chi_{\uparrow\downarrow}-\chi_{\downarrow\uparrow}+\chi_{\downarrow\downarrow});\qquad
\chi_{\sigma\sigma'}(\mbf r,t;\mbf r',t')=-\ii\theta(t-t')\avg{[\hat n_\sigma(\mbf r,t),\hat n_{\sigma'}(\mbf r',t')]}.
\end{gather}

The magnetic susceptibility is thus a combination of spin-resolved density--density susceptibilities. We switch to Fourier transform and write
\begin{equation}
\chi_{\sigma\sigma'}(\mbf r,t;\mbf r',t')=\frac1{\mathcal V}\sum_\mbf{q}e^{\ii \mbf{q}\cdot(\mbf r-\mbf r')}\int\frac{\dd\omega}{2\pi}\,e^{-\ii\omega(t-t')}\chi_{\sigma\sigma'}(\mbf{q},\omega).\end{equation}

For a nonuniform field
\begin{equation}M(\mbf{q},\omega)=\chi_M(\mbf{q},\omega)B(\mbf{q},\omega),\quad\chi_M(\mbf{q},\omega)=-\mu_B^2[\chi_{\uparrow\uparrow}(\mbf{q},\omega)+\chi_{\downarrow\downarrow}(\mbf{q},\omega)-\chi_{\uparrow\downarrow}(\mbf{q},\omega)-\chi_{\downarrow\uparrow}(\mbf{q},\omega)].
\end{equation}
For a uniform field we have $M(\omega)=\chi_M(\omega)B(\omega)$, where \begin{equation}
  \chi_M(\omega)=-\mu_B^2[\chi_{\uparrow\uparrow}(\mbf{q}=0,\omega)+\chi_{\downarrow\downarrow}(\mbf{q}=0,\omega)-\chi_{\uparrow\downarrow}(\mbf{q}=0,\omega)-\chi_{\downarrow\uparrow}(\mbf{q}=0,\omega)].
\end{equation}

\section{Susceptibility for external fermionic bath}\label{App:External bath}

In this section we detail the calculation of the susceptibility when the electron gas is coupled to a fermionic bath.

We start from the expressions given in the main text
\begin{gather}
\chi_{\sigma\sigma'}(\mbf q,\omega)=-\frac\ii2\frac1{\mathcal V}\sum_k\int\frac{\dd\varepsilon}{2\pi}\big[G_{\sigma\sigma'}^R(\mbf{k},\varepsilon)G_{\sigma'\sigma}^K(\mbf{\mbf{k}}-\mbf{q},\varepsilon-\omega)+G_{\sigma\sigma'}^K(\mbf{k},\varepsilon)G_{\sigma'\sigma}^A(\mbf{\mbf{k}}-\mbf{q},\varepsilon-\omega)\big],\\
G_{\sigma\sigma'}^K(r,t;r',t')=-\ii\avg{[\hat\psi_\sigma(r,t),\hat\psi_{\sigma'}^\dagger(r',t')]},\qquad
G_{\sigma\sigma'}^K(\mbf{k},\varepsilon)=[1-2f(\varepsilon)][G_{\sigma\sigma'}^R(\mbf{k},\varepsilon)-G_{\sigma\sigma'}^A(\mbf{k},\varepsilon)].
\end{gather}

We assume that thermalization is induced by an external fermionic bath at inverse temperature $\beta$, which interacts with the electron gas via hopping terms. We write, in momentum space
\begin{equation}
\hat{H}_{\rm{tot}}=\sum_{\kk\sigma}\varepsilon_\kk\hat c^\dagger_{\kk\sigma}\hat c_{\kk\sigma}+\sum_{\kk,\sigma,a}\varepsilon^{(\rm b)}_{\kk a} \hat b^\dagger_{\kk\sigma a}\hat b_{\kk\sigma a} +\sum_{\kk,\sigma,a} \bigl(t_{\kk a}\hat c^\dagger_{\kk\sigma}\hat b_{\kk\sigma a} +\mathrm{h.c.}\bigr)
\end{equation}
where $\hat b_{\kk\sigma a}$ is the annihilation operator of a bath mode $a$ with momentum $\kk$ and spin $\sigma$, whose energy is $\varepsilon^{(\rm b)}_{\kk a}$. We assume the bath is held at inverse temperature $\beta$, so that its equilibrium distribution is $f(\varepsilon)$. The coupling $t_{\kk a}$ is assumed spin-independent and spin diagonal. Integrating out, the bath degrees of freedom we find
\begin{gather}
(G^{R/A})^{-1}(\kk,\varepsilon)=\varepsilon-\varepsilon_\kk-\Sigma^{R/A}(\kk,\varepsilon); \quad (G^{-1})^K(\kk,\varepsilon)=-\Sigma^K(\kk,\varepsilon),
\end{gather}
where $\Sigma$ is the self-energy due to the interaction with the bath, given by
\begin{equation}
\Sigma^{R/A}(\kk,\varepsilon)=\sum_a\frac{|t_{\kk a}|^2}{\varepsilon-\varepsilon^{(\rm b)}_{\kk a}\pm\ii\eta}; \qquad \Sigma^K(\kk,\varepsilon)=[1-2f(\varepsilon)][\Sigma^R(\kk,\varepsilon)-\Sigma^A(\kk,\varepsilon)].
\end{equation}

We now take use $\eta\rightarrow0^+$, to calculate the real and imaginary part of $\Sigma^R$. The real part is proportional to $\sim|t_{\kk a}|^2$ and produces a negligible shift of the single-particle energy $\varepsilon_{\kk}$. For the imaginary part we find
\begin{equation}
\Sigma^R(\kk,\varepsilon)=-\ii\pi\sum_a\delta(\varepsilon-\varepsilon^{(\rm b)}_{\kk a})|t_{\kk a}|^2\equiv-\ii\kappa;\qquad \Sigma^A(\kk,\varepsilon)=\ii\kappa; \qquad \Sigma^K(\kk,\varepsilon)=-2\ii\kappa[1-2f(\varepsilon)].
\end{equation}
Here we have assumed that the band of the bath modes is flat enough and the $\kk$ dependence is negligible, so that the self-energy is constant (within the range of energies relevant to our system).

We now use that $G^K=-G^R(G^{-1})^KG^A$ \cite{Kamenev01052009} and find
\begin{equation}G_{\sigma\sigma'}^{R/A}=\frac{\delta_{\sigma\sigma'}}{\varepsilon-\varepsilon_{\mbf k}\pm\ii\kappa},\qquad G_{\sigma\sigma'}^K=\delta_{\sigma\sigma'}[1-2f(\varepsilon)]\frac{-2\ii\kappa}{(\varepsilon-\varepsilon_{\mbf k})^2+\kappa^2},
\end{equation}
from which we correctly retrieve that $G^K=[1-2f(\varepsilon)](G^R-G^A)$. Since the Green's functions remain spin diagonal and produce the same contribution from both spin sectors, we can write
\begin{equation}\label{Eq:AppChiM}
\chi_M(\omega)=\ii\mu_B^2\frac1{\mathcal V}\sum_\mbf{k}\int\frac{\dd\varepsilon}{2\pi}\big[G^R(\mbf{k},\varepsilon)G^K(\mbf{\mbf{k}},\varepsilon-\omega)+G^K(\mbf{k},\varepsilon)G^A(\mbf{\mbf{k}},\varepsilon-\omega)\big]
\end{equation}

In order to calculate the adiabatic susceptibility, we first take the $\kappa\rightarrow0$ limit, which gives $G^{R/A}=(\varepsilon-\varepsilon_{\mbf k}\pm\ii\eta)^{-1}$ and $G^K=-2\pi\ii(1-2f(\varepsilon)\delta(\varepsilon-\varepsilon_{\mbf k})$ so that
\begin{equation}
\begin{aligned}
\chi_M^S&=\lim_{\omega\rightarrow0}\ii\frac{\mu_B^2}{\mathcal{V}}\sum_{\mbf k}\int\dd\varepsilon(-\ii)\left[\frac{1-2f(\varepsilon-\omega)}{\varepsilon-\varepsilon_{\mbf k}+\ii\eta}\delta(\varepsilon-\omega-\varepsilon_{\mbf k})+\frac{1-2f(\varepsilon)}{\varepsilon-\omega-\varepsilon_{\mbf k}+\ii\eta}\delta(\varepsilon-\varepsilon_{\mbf k})\right]=\\
&=\lim_{\omega\rightarrow0}\frac{\mu_B^2}{\mathcal{V}}\sum_{\mbf k}\left[\frac{1-2f(\varepsilon_{\mbf k})}{\omega+\ii\eta}+\frac{1-2f(\varepsilon_{\mbf k})}{-\omega-\ii\eta}\right]=0.
\end{aligned}
\end{equation}

In order to calculate the isothermal susceptibility, we have to take $\omega\to0$ before $\kappa\to0$:
\begin{gather}
\begin{gathered}
\chi_M^T=\lim_{\kappa\to0}\frac{\ii\mu_B^2}{\mathcal V}\sum_{\mbf k}\int\frac{\dd\varepsilon}{2\pi}[G^R(\mbf k,\varepsilon)G^K(\mbf k,\varepsilon)+G^K(\mbf k,\varepsilon)G^A(\mbf k,\varepsilon)]=\lim_{\kappa\to0}\frac{\ii\mu_B^2}{\mathcal V}\sum_{\mbf k}I(\mbf{k}),\\
I(\mbf{k})=\int\frac{\dd\varepsilon}{2\pi}(-2\ii\kappa)[1-2f(\varepsilon)]\left[\frac1{(\varepsilon-\varepsilon_{\mbf k}+\ii\kappa)((\varepsilon-\varepsilon_{\mbf k})^2+\kappa^2)}+\frac1{((\varepsilon-\varepsilon_{\mbf k})^2+\kappa^2)(\varepsilon-\varepsilon_{\mbf k}-\ii\kappa)}\right]\\
=-4\ii\kappa\int\frac{\dd\varepsilon}{2\pi}\frac{(\varepsilon-\varepsilon_k)[1-2f(\varepsilon)]}{[(\varepsilon-\varepsilon_k)^2+\kappa^2]^2}
=2\ii\kappa\int\frac{\dd\varepsilon}{2\pi}[1-2f(\varepsilon)]\frac\partial{\partial\varepsilon}\frac1{(\varepsilon-\varepsilon_k)^2+\kappa^2}
=4\ii\kappa\int\frac{\dd\varepsilon}{2\pi}\frac{\partial f/\partial\varepsilon}{(\varepsilon-\varepsilon_\mbf k)^2+\kappa^2};
\end{gathered}\\
\chi_M^T=\lim_{\kappa\to0}\frac{\mu_B^2}{\mathcal V}\sum_{\mbf k\sigma}\int\frac{\dd\varepsilon}{2\pi}\frac{\partial f}{\partial\varepsilon}\frac{-2\kappa}{(\varepsilon-\varepsilon_\mbf k)^2+\kappa^2}=\frac{\mu_B^2}{\mathcal V}\sum_{\mbf k\sigma}\left(-\frac{\partial f}{\partial\varepsilon_{\mbf k}}\right),
\end{gather}
where we have used that $\lim_{\kappa\rightarrow0}\frac{\kappa}{(\varepsilon-\varepsilon_\mbf k)^2+\kappa^2}=\pi\delta(\varepsilon-\varepsilon_\kk)$. At low temperatures $\partial f/\partial\varepsilon_{\mbf k}\approx-\delta(\varepsilon_{\mbf k})$ so that
\begin{equation}\chi_M^T=\frac{\mu_B^2}{\mathcal V}\sum_{\sigma,\mbf{k}}\delta(\varepsilon_F-\mbf k^2/2m)=\mu_B^2\mathcal N(\varepsilon_F).
\end{equation}

At low temperatures, we can actually derive an explicit expression for $\chi_M(\omega,\kappa)$. We start from Eq. \eqref{Eq:AppChiM} and replace $\varepsilon\rightarrow\varepsilon+\omega$ in the first term so that we find
\begin{gather}
\chi_M(\omega,\kappa)=\frac{\ii\mu_B^2}{\mathcal V}\sum_\mbf{k}\int\frac{\dd\varepsilon}{2\pi}\big[G^R(\mbf{k},\varepsilon+\omega)G^K(\mbf{\mbf{k}},\varepsilon)+G^K(\mbf{k},\varepsilon)G^A(\mbf{\mbf{k}},\varepsilon-\omega)\big]=\\
=\frac{\ii\mu_B^2}{\mathcal V}\sum_\mbf{k}\int\frac{\dd\varepsilon}{2\pi}(1-2f(\varepsilon))\frac{-2\ii\kappa}{(\varepsilon-\varepsilon_\mbf{k})^2+\kappa^2}\left[\frac{1}{\varepsilon-\varepsilon_\mbf{k}+\omega+\ii\kappa}+\frac{1}{\varepsilon-\varepsilon_\mbf{k}-\omega-\ii\kappa}\right]=\\
=\frac{\ii\mu_B^2}{\mathcal V}\sum_\mbf{k}\int\frac{\dd\varepsilon}{2\pi}(1-2f(\varepsilon))\frac{-2\ii\kappa}{(\varepsilon-\varepsilon_\mbf{k})^2+\kappa^2}\frac{2(\varepsilon-\varepsilon_\kk)}{(\varepsilon-\varepsilon_\mbf{k})^2-(\omega+\ii\kappa)^2}=\frac{\ii\mu_B^2}{\mathcal V}\sum_\mbf{k}J(\kk,\omega);\\
  J(\kk,\omega)=\int\frac{\dd\varepsilon}{2\pi}\,(1-2f(\varepsilon))\,F(\varepsilon-\varepsilon_\kk),\qquad F(u)\equiv\frac{-4\ii\kappa\,u}{(u^2+\kappa^2)(u^2-z^2)}; \qquad z\equiv\omega+\ii\kappa.
\label{eq:F}
\end{gather}

We define $\cF(u)\equiv\int_{-\infty}^{u}F(u')\,du'$. Since $F$ is odd and integrable, $\cF(+\infty)=\int_{-\infty}^{\infty}F=0$, so that integrating Eq.~\eqref{eq:F} by parts we find,
\begin{equation}
  \int \dd\varepsilon\,(1-2f(\varepsilon))F(\varepsilon-\varepsilon_\kk)
  =\Big[(1-2f(\varepsilon))\,\cF(\varepsilon-\varepsilon_\kk)\Big]_{-\infty}^{+\infty}
   +2\int\dd\varepsilon\,f'(\varepsilon)\,\cF(\varepsilon-\varepsilon_\kk)=2\int\dd\varepsilon\,f'(\varepsilon)\,\cF(\varepsilon-\varepsilon_\kk).
\end{equation}
The boundary term vanishes because $h(\pm\infty)=1-2f(\pm\infty)=\pm1$ is bounded and $\cF(\pm\infty)=0$:
\begin{equation}
  J(\kk,\omega)=\frac{1}{\pi}\int d\varepsilon\,f'(\varepsilon)\,\cF(\varepsilon-\varepsilon_\kk).
  \label{eq:step3}
\end{equation}

The derivative $f'$ restricts the energy integral to the Fermi level. We introduce an integral over $\xi$ to write the sum over $\kk$ in terms of the density of states:
\begin{equation}
  \frac1{\mathcal{V}}\sum_\kk J(\kk,\omega)=\frac1{2\pi}\int d\varepsilon\,f'(\varepsilon)\int_{-\infty}^{\infty}d\xi\,\cF(\varepsilon-\xi)\frac1{\mathcal{V}}\sum_{\kk\sigma}\delta(\xi-\varepsilon_\kk)=\frac1{2\pi}\int d\varepsilon\,f'(\varepsilon)\int_{-\infty}^{\infty}d\xi\,\cF(\varepsilon-\xi)\mathcal{N}(\varepsilon_F+\xi),
  \label{eq:4a}
\end{equation}
where $\mathcal{N}(\xi+\varepsilon_F)=\frac1{\mathcal{V}}\sum_{\kk\sigma}\delta(\xi-\varepsilon_\kk)$. We now assume that the temperature is very low so that $f'(\varepsilon)\approx-\delta(\varepsilon)$. Moreover, the function $\mathcal{F}(\xi)$ probes the energies around $\xi\sim0$, with a width of order $\text{max}(\omega,\kappa)\ll\varepsilon_F$. Thus we can approximate $\mathcal{N}(\varepsilon_F+\xi)\approx\mathcal{N}(\varepsilon_F)$ and write
\begin{equation}
\chi_M(\omega,\kappa)\approx-\ii\mu_B^2\frac1{2\pi}\int_{-\infty}^{\infty}\dd\xi\,\cF(-\xi)\mathcal{N}(\varepsilon_F)=-\ii\chi_P\frac1{2\pi}\int_{-\infty}^{\infty}\dd u\,\cF(u).
\end{equation}

We can integrate by parts and write
\begin{equation}
 \int_{-\infty}^{\infty}\dd u\,\cF(u)=\Big[u\,\cF(u)\Big]_{-\infty}^{+\infty}-\int_{-\infty}^{\infty}\dd u\,u\,F(u)
     =-\int_{-\infty}^{\infty}\dd u\,u\,F(u)=4\ii\kappa\int_{-\infty}^{\infty}\dd u\frac{u^2}{[(u^2+\kappa^2)(u^2-z^2)]}.
\end{equation}

We use that 
\begin{gather}
  \frac{u^2}{(u^2+\kappa^2)(u^2-z^2)}
  =\frac{1}{\kappa^2+z^2}\left[\frac{\kappa^2}{u^2+\kappa^2}+\frac{z^2}{u^2-z^2}\right];\\
  \int_{-\infty}^{\infty}\frac{du}{u^2+\kappa^2}=\frac{\pi}{\kappa};\qquad
  \int_{-\infty}^{\infty}\frac{du}{u^2-z^2}
  =\,\mathop{\mathrm{Res}}_{u=z}\frac{2\pi \ii}{(u-z)(u+z)}
  =2\pi \ii\,\frac{1}{2z}=\frac{\ii\pi}{z}.
\end{gather}

Therefore we find
\begin{gather}
  \int_{-\infty}^{\infty}\dd u\frac{u^2}{[(u^2+\kappa^2)(u^2-z^2)]}=\frac{\pi(\kappa+\ii z)}{\kappa^2+z^2}=\frac{\ii\pi}{z+\ii\kappa}=\frac{\ii\pi}{\omega+2\ii\kappa};
  \label{eq:K}\\
\chi_M(\omega,\kappa)\approx\chi_P\frac{2\kappa}{\pi}\frac{\ii\pi}{\omega+2\ii\kappa}=\chi_P\frac{2\ii\kappa}{\omega+2\ii\kappa}=\chi_P\frac{2\kappa}{2\kappa-\ii\omega}.
\end{gather}

\section{Spin-flip Hamiltonian}\label{App:spin flip}

In this section we study the case in which  the Hamiltonian contains a term that couples the two spin sectors and breaks the conservation of $\hat S_z$. We consider a simple spin-flipping term:
\begin{equation}
\hat H_{\rm{sf}}=\kappa\sum_{\mbf k}(\hat c_{\mbf k\uparrow}^\dagger \hat c_{\mbf k\downarrow}+\hat c_{\mbf k\downarrow}^\dagger \hat c_{\mbf k\uparrow}),\qquad\hat H_0=\sum_{\mbf k}\psi_{\mbf k}^\dagger\begin{pmatrix}\varepsilon_{\mbf k}&\kappa\\\kappa&\varepsilon_{\mbf k}\end{pmatrix}\psi_{\mbf k},\quad\psi_{\mbf k}=\begin{pmatrix}\hat c_{\mbf k\uparrow}\\\hat c_{\mbf k\downarrow}\end{pmatrix}.
\end{equation}

The Green's function is a matrix, and can be obtained by inverting $G^{-1}(\mbf k,\varepsilon)=\varepsilon-\varepsilon_{\mbf k}-\kappa\tau^x$ with $\tau^x=\begin{pmatrix}0&1\\1&0\end{pmatrix}$:
\begin{align}
G_{\uparrow\uparrow}^{R/A}(\mbf k,\varepsilon)&=\frac{\varepsilon-\varepsilon_{\mbf k}\pm\ii\eta}{(\varepsilon-\varepsilon_{\mbf k}-\kappa\pm\ii\eta)(\varepsilon-\varepsilon_{\mbf k}+\kappa\pm\ii\eta)}=\frac12\left[\frac1{\varepsilon-\varepsilon_{\mbf k}-\kappa\pm\ii\eta}+\frac1{\varepsilon-\varepsilon_{\mbf k}+\kappa\pm\ii\eta}\right]=G_{\downarrow\downarrow}^{R/A},\\
G_{\uparrow\downarrow}^{R/A}(\mbf k,\varepsilon)&=\frac\kappa{(\varepsilon-\varepsilon_{\mbf k}-\kappa\pm\ii\eta)(\varepsilon-\varepsilon_{\mbf k}+\kappa\pm\ii\eta)}=\frac12\left[\frac1{\varepsilon-\varepsilon_{\mbf k}-\kappa\pm\ii\eta}-\frac1{\varepsilon-\varepsilon_{\mbf k}+\kappa\pm\ii\eta}\right]=G_{\downarrow\uparrow}^{R/A},\\
G_{\uparrow\uparrow}^K(k,\varepsilon)&=-\ii\pi[1-2f_D(\varepsilon)][\delta(\varepsilon-\varepsilon_{\mbf k}-\kappa)+\delta(\varepsilon-\varepsilon_{\mbf k}+\kappa)],\\
G_{\uparrow\downarrow}^K(k,\varepsilon)&=-\ii\pi[1-2f_D(\varepsilon)][\delta(\varepsilon-\varepsilon_{\mbf k}-\kappa)-\delta(\varepsilon-\varepsilon_{\mbf k}+\kappa)].
\end{align}

The integral we need to calculate to obtain $\chi_{\sigma\sigma'}$ is
\begin{equation}
I_{\sigma\sigma'}(k)=\int\frac{\dd\varepsilon}{2\pi}[G_{\sigma\sigma'}^R(\mbf k,\varepsilon)G_{\sigma'\sigma}^K(\mbf k-\mbf{q},\varepsilon-\omega)+G_{\sigma\sigma'}^K(\mbf k,\varepsilon)G_{\sigma'\sigma}^A(\mbf k-\mbf{q},\varepsilon-\omega)].
\end{equation}

For compactness let $D=\omega+\varepsilon_{k-\mbf{q}}-\varepsilon_{\mbf k}+\ii\eta$. After performing the integration of the delta functions in $I_{\uparrow\uparrow}$ we obtain eight terms:
\begin{gather}
\begin{aligned}
I_{\uparrow\uparrow}=-\frac\ii2\int\dd\varepsilon[&G^R_{\uparrow\uparrow}(\mbf k,\varepsilon)(1-2f(\varepsilon-\omega))[\delta(\varepsilon-\omega-\varepsilon_{\mbf k-\mbf q}-\kappa)+\delta(\varepsilon-\omega-\varepsilon_{\mbf k-\mbf q}+\kappa)]+\\
&(1-2f(\varepsilon))[\delta(\varepsilon-\varepsilon_{\mbf k}-\kappa)+\delta(\varepsilon-\varepsilon_{\mbf k}+\kappa)]G^A_{\uparrow\uparrow}(\mbf k-\mbf q,\varepsilon-\omega)];
\end{aligned}\\
\begin{aligned}
I_{\uparrow\uparrow}=&-\frac\ii4\left[\frac{1-2f(\varepsilon_{k-\mbf{q}}+\kappa)}D+\frac{1-2f(\varepsilon_{\kk-\mbf{q}}+\kappa)}{D+2\kappa}+\frac{1-2f(\varepsilon_{\kk-\mbf{q}}-\kappa)}{D-2\kappa}+\frac{1-2f(\varepsilon_{\kk-\mbf{q}}-\kappa)}D\right]\\
&-\frac\ii4\left[\frac{1-2f(\varepsilon_{\mbf k}+\kappa)}{-D}+\frac{1-2f(\varepsilon_{\mbf k}+\kappa)}{-D+2\kappa}+\frac{1-2f(\varepsilon_{\mbf k}-\kappa)}{-D-2\kappa}+\frac{1-2f(\varepsilon_{\mbf k}-\kappa)}{-D}\right]\\
=&-\frac\ii{2D}[f(\varepsilon_{\mbf k}+\kappa)-f(\varepsilon_{\mbf k-\mbf{q}}+\kappa)+f(\varepsilon_{\mbf k}-\kappa)-f(\varepsilon_{\mbf k-\mbf{q}}-\kappa)]\\
&-\frac\ii2\frac{f(\varepsilon_{\mbf k}-\kappa)-f(\varepsilon_{\mbf k-\mbf{q}}+\kappa)}{D+2\kappa}-\frac\ii2\frac{f(\varepsilon_{\mbf k}+\kappa)-f(\varepsilon_{\mbf k-\mbf{q}}-\kappa)}{D-2\kappa}.
\end{aligned}
\end{gather}

Similarly
\begin{gather}
\begin{aligned}
I_{\uparrow\downarrow}=&-\frac\ii{2D}[f(\varepsilon_{\mbf k}+\kappa)-f(\varepsilon_{k-\mbf{q}}+\kappa)+f(\varepsilon_{\mbf k}-\kappa)-f(\varepsilon_{\kk-\mbf{q}}-\kappa)]-\\
&-\frac\ii2\frac{f(\varepsilon_{\kk-\mbf{q}}+\kappa)-f(\varepsilon_{\mbf k}-\kappa)}{D+2\kappa}-\frac\ii2\frac{f(\varepsilon_{\kk-\mbf{q}}-\kappa)-f(\varepsilon_{\mbf k}+\kappa)}{D-2\kappa};
\end{aligned}\\
I_{\uparrow\uparrow}-I_{\uparrow\downarrow}=-\ii\left[\frac{f(\varepsilon_{\mbf k}-\kappa)-f(\varepsilon_{k-\mbf{q}}+\kappa)}{D+2\kappa}+\frac{f(\varepsilon_{\mbf k}+\kappa)-f(\varepsilon_{\kk-\mbf{q}}-\kappa)}{D-2\kappa}\right].
\end{gather}

The magnetic susceptibility is then (taking $\eta\rightarrow0$)
\begin{equation}
\begin{gathered}
\begin{aligned}
\chi_M(\mbf q,\omega,\kappa)&=\frac{\mu_B^2}2\frac1{\mathcal V}\sum_{\mbf k\sigma}\left[\frac{f(\varepsilon_{\mbf k}-\kappa)-f(\varepsilon_{\mbf k-\mbf q}+\kappa)}{\omega+\varepsilon_{\mbf k-\mbf q}-\varepsilon_{\mbf k}+2\kappa+\ii\eta}+\frac{f(\varepsilon_{\mbf k}+\kappa)-f(\varepsilon_{\mbf k-\mbf q}-\kappa)}{\omega+\varepsilon_{\mbf k-\mbf q}-\varepsilon_{\mbf k}-2\kappa+\ii\eta}\right];\\
\chi_M(\mbf q=0,\omega,\kappa)&=\frac{\mu_B^2}2\frac1{\mathcal V}\sum_{\mbf k\sigma}[f(\varepsilon_{\mbf k}-\kappa)-f(\varepsilon_{\mbf k}+\kappa)]\frac{-4\kappa}{(\omega+\ii\eta)^2-4\kappa^2};
\end{aligned}\\
\chi_M(\mbf q=0,\omega,0)=0\quad\Rightarrow\quad\chi_M^S=\lim_{\omega\to0}\lim_{\kappa\to0}\chi_M(0,\omega,\kappa)=0;\\
\chi_M^T=\lim_{\kappa\to0}\lim_{\omega\to0}\chi_M(0,\omega,\kappa)=\lim_{\kappa\to0}\frac{\mu_B^2}2\frac1{\mathcal V}\sum_{\mbf k\sigma}\frac{f(\varepsilon_{\mbf k}-\kappa)-f(\varepsilon_{\mbf k}+\kappa)}\kappa
=\mu_B^2\frac1{\mathcal V}\sum_{\mbf k\sigma}\left(-\frac{\partial f}{\partial\varepsilon_{\mbf k}}\right)=\chi_P.
\end{gathered}
\end{equation}

We thus recover the same results obtained when the system is coupled to a fermionic bath, even though $\hat H_{\rm{sf}}$ does not induce any thermalization. These results can prove useful for systems used in adiabatic annealing, such as the transverse Ising model \cite{kadowakiQuantumAnnealingTransverse1998,dasColloquiumQuantumAnnealing2008,shevchenkoLandauZenerStuckelbergInterferometry2010,dicksonThermallyAssistedQuantum2013,aminSearchingQuantumSpeedup2015,marshallPowerPausingAdvancing2019}. 

For $\mbf q\neq0$
\begin{gather}
\chi_M^T(\mbf q)=\lim_{\kappa\to0}\frac{\mu_B^2}{2\mathcal V}\sum_{\mbf k\sigma}\left[\frac{f(\varepsilon_{\mbf k}-\kappa)-f(\varepsilon_{\mbf k-\mbf q}+\kappa)}{\varepsilon_{\mbf k-\mbf q}-\varepsilon_{\mbf k}+2\kappa+\ii\eta}+\frac{f(\varepsilon_{\mbf k}+\kappa)-f(\varepsilon_{\mbf k-\mbf q}-\kappa)}{\varepsilon_{\mbf k-\mbf q}-\varepsilon_{\mbf k}-2\kappa+\ii\eta}\right]=\frac{\mu_B^2}{\mathcal V}\sum_{\mbf k\sigma}\frac{f(\varepsilon_{\mbf k})-f(\varepsilon_{\mbf k-\mbf q})}{\varepsilon_{\mbf k-\mbf q}-\varepsilon_{\mbf k}+\ii\eta};\\
\chi_M^S(\mbf q)=\lim_{\omega\to0}\lim_{\kappa\to0}\chi_M(\mbf q,\omega,\kappa)=\frac{\mu_B^2}{\mathcal V}\sum_{\mbf k\sigma}\frac{f(\varepsilon_{\mbf k})-f(\varepsilon_{\mbf k-\mbf q})}{\varepsilon_{\mbf k-\mbf q}-\varepsilon_{\mbf k}+\ii\eta}=\chi_M^T(\mbf q);\\
\lim_{\mbf q\to0}\chi_M^{T/S}(\mbf q)=\frac{\mu_B^2}{\mathcal V}\sum_{\mbf k\sigma}\left(-\frac{\partial f}{\partial\varepsilon_{\mbf k}}\right)=\chi_M^T=\chi_P.
\end{gather}

\section{A classical analogue}
\label{App:class}
In this section we study a simple classical system which exhibits a different response for the adiabatic and isothermal regimes.

\begin{figure}
    \centering
    \includegraphics[width=0.6\linewidth]{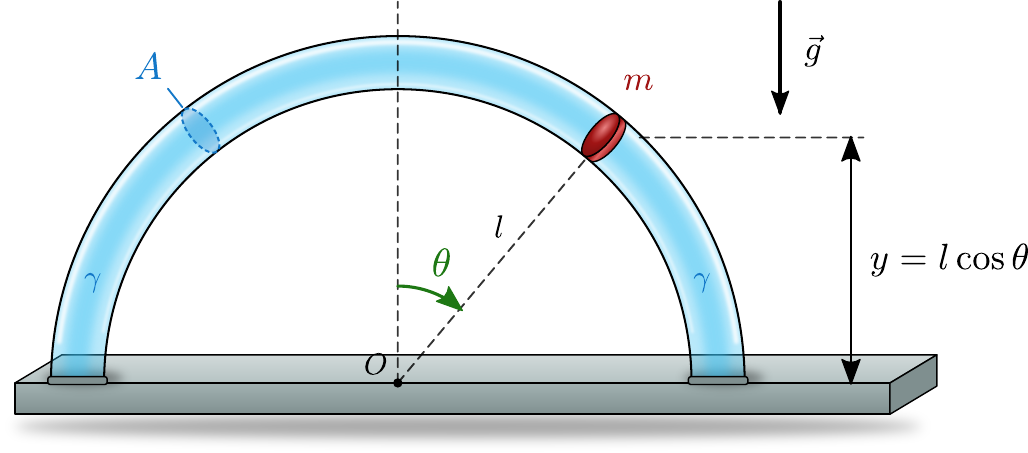}
    \caption{Sketch of the classical system under considerations. A mass $m$ divides a semicircular tube of section $A$ and radius $l$ in two parts filled with an ideal gas. This image was AI generated with Claude-Opus5.5.}
    \label{fig:classical}
\end{figure}

We consider a circular tube with cross-section $A$, shaped like a semicircle of radius $l$, filled with an ideal gas with an adiabatic coefficient $\gamma$. Inside the tube there is a mass $m$ of negligible thickness and a cross-section equal to that of the tube; the mass can move freely and without friction within the tube. The semicircle is placed vertically on a table, so that the position of the mass is determined by the angle $\theta$ it makes with the vertical, and its vertical coordinate is given by $y=l\cos\theta$, see Fig. \ref{fig:classical}.

Initially, the mass divides the tube into two halves, each containing $n$ moles of gas at temperature $T_0$. We wish to analyze the motion of the mass in the adiabatic and isothermal regimes, in which the system is kept isolated or in contact with a thermostat at temperature $T_0$, respectively.

\subsection{Adiabatic regime}

Suppose the tube has insulating walls, so that the gas inside each of the two halves of the tube defined by the mass transforms adiabatically. The pressure of the two halves transforms as $PV^\gamma=P_0V_0^\gamma$, where $P_0=nRT_0/V_0$ and $V_0=\frac\pi2Al$. Thus
\begin{equation}
P_{R/L}(\theta)=P_0(V_0/V)^\gamma=\frac{nRT_0}{\frac{\pi}{2}Al}\left(1\mp\frac{2\theta}{\pi}\right)^{-\gamma}   
\end{equation}

where the sign $-$ and $+$ are associated with the right and left halves, respectively.
The total force due to the pressure difference acting on the mass is then $F=-(P_R-P_L)A$. We thus find the equation of motion
\begin{gather}
    ml\ddot{\theta}=mg\sin\theta-\frac{nRT_0}{\frac\pi2l}\left(\frac{1}{(1-2\theta/\pi)^\gamma}-\frac{1}{(1+2\theta/\pi)^\gamma}\right).
\end{gather}

We want to study the stability of the equilibrium position $\theta=0$ in the absence of external forces. We expand up to linear terms and find
\begin{gather}
    ml\ddot{\theta}\approx mg\theta-\frac{8nRT_0}{\pi^2l}\gamma\theta=-ml(\omega_S^2-g/l)\theta; \qquad \omega_S^2\equiv\gamma\frac{8nRT_0}{\pi^2ml^2}
\end{gather}

Where $\omega_S$ is the adiabatic response frequency of the gas. The equilibrium position $\theta=0$ is stable if $\omega_S^2>g/l$, in which case the system undergoes small oscillations with frequency $\omega=\sqrt{\omega_S^2-g/l}$.

\subsection{Isothermal regime}

Let us assume now that the walls of the tube can conduct heat in a very efficient way, so that the temperature of the gas is always fixed to $T_0$. In this case the pressure transforms according to isothermal law $PV=P_0V_0$ and the equation of motion becomes
\begin{gather}
    ml\ddot{\theta}=mg\sin\theta-\frac{nRT_0}{\frac\pi2l}\left(\frac{1}{1-2\theta/\pi}-\frac{1}{1+2\theta/\pi}\right).
\end{gather}

In this case the small oscillations expansion becomes
\begin{gather}
    ml\ddot{\theta}\approx mg\theta-\frac{8nRT_0}{\pi^2l}\theta=-ml(\omega_T^2-g/l)\theta; \qquad \omega_T^2\equiv\frac{8nRT_0}{\pi^2ml^2}.
\end{gather}

where the isothermal response frequency is $\omega_T=\omega_S/\sqrt{\gamma}$ and the stability condition is $\omega_T^2>g/l$. 

Since $\omega_T<\omega_S$, there is a regime of parameters for which the adiabatic regime exhibits stable oscillations around the vertical, while in the isothermal regime the vertical position is unstable and the mass is in equilibrium at $\theta\neq0$.

\end{document}